\documentstyle[preprint,aps]{revtex}

\begin{document}
\setcounter{page}{1}

\draft

\title{
 Mass hierarchy from compositeness hierarchy
in supersymmetric gauge theory
}

\author{
 M. Hayakawa \footnotemark[0] 
}
\footnotetext[0]
{ 
JSPS Research Fellow \\
Electronic address: hayakawa@theory.kek.jp
}

\address{
Division of Theoretical Physics, KEK, \\
Tsukuba, Ibaraki 305, Japan
}

\preprint{\begin{minipage}{4cm}
           KEK-TH-516 \\
           April 1997 \\
          \end{minipage}
         }

\date{\today}

\maketitle

\begin{abstract}
  The possibility that the mass hierarchy
is intimately associated with the compositeness level
of the matters is proposed in supersymmetric gauge theory.
  This implies, for instance, that
the preons constituting top quark consists of the ``prepreons''
binded by the same gauge force making the charm quark
out of another preons.
  The exemplifying toy model illustrates how
the hierarchy in the yukawa coupling constants
in the up-quark sector
is generated from the underlying gauge dynamics.
  It is also indicated that
the incorporation of down-type quarks as elementary objects
leads to unpleasant results generically.
 Thus all the quarks as well as
the leptons must also be regarded as composite
in the present approach.
\end{abstract}

\pacs{ }

\narrowtext
%

\section{Introduction}
\label{sec:intro}

 The hierarchic structure observed in the masses
among the different generations of quarks and leptons
is one of the questions
that cannot be addressed to within the standard model.
 The yukawa sector of the standard model
consists of many coupling constants
which provide a necessary and sufficient number of parameters
to reproduce the masses and mixing of quarks
as well as (weak) CP phase at the present experimental stage.
 In particular there cannot be found any reason
why such different orders of numerals
appear in the yukawa coupling constants.

 In a recent few years
various composite models of quarks (and leptons)
have been proposed so far \cite{Strassler,NS,LM},
using the knowledge on the nonperturbative
aspects of supersymmetric gauge theory
\cite{Seiberg}.
 Those models utilize the dynamics of
a series of the asymptotically free gauge theory
\cite{Csaki};
confinement without spontaneous breakdown
of chiral symmetry.
 It generates dynamically the nonperturbative terms
which accommodate the yukawa interactions of
the quarks (and leptons).

 In this note
we would like to propose another class of composite models
to access to the mass hierarchy problem
by using the same type of supersymmetric gauge dynamics.
 That is, the hierarchy of masses
may have its origin intimately associated with
the {\it level} of the compositeness
of each generation of the matters.
 It may open up the potential possibility
to construct the models possessing more and more predictive power.

 The model of Ref. \cite{NS}
prepares three types of preons,
each of which is subject to
asymptotically free SU(2) gauge interaction
but characterized by the dynamical scales different from each other.
 Thus, there,
the hierarchy of the masses in the up-type quarks,
for instance,
is traced back to
the hierarchy of the scales of
the SU(2) gauge dynamics;
the repetition of the same structure for confining forces
characterized by three different scales.

 The matter structure examined here is as follows. 
 For simplicity let us take our attention to
the last two generations of the up-type quarks,
i.e., charm and top quarks,
and consider their respective potential compositeness.
 The charm quark is assumed to consist
of the preons subject to the confining interactions
of the gauge group ${\rm G}_2$.
 The same force ${\rm G}_2$ binds the ``prepreons''
constituting the preons,
which will be binded
by asymptotically free ${\rm G}_1$ gauge force
at lower energy scale to form top quark.
 Thus as long as we restrict to those two species,
the top quarks is considered to be the composite of
the composite preons
while the charm quark the composite of the preons.
 To uncover the requisite pattern of hierarchy,
Higgs doublet giving the up-type quarks
should also be composite with the same level of compositeness
as top quark.
 The same consideration
can also be generalized in the case of the three generations.

 The next section gives a toy model
how such a structure give rise to the hierarchy
in the masses of quarks,
i.e., the hierarchy in the strength of yukawa coupling,
those of which arise from the underlying gauge theory,
especially sticking to the two generation.
 This model itself is far from reality
but illustrates
the main ingredients of the models
based on the idea proposed here.

 The last section will be devoted to
the discussion and conclusion.

\section{Example of prepreon model}
\label{sec:prepreon}

 The simplest example of the models realizing the idea
described in the previous section
is inferred from the use
of the infrared dynamics of
the supersymmetric ${\rm SP}(N_c)$ gauge theory
with $N_f$ = $N_c + 2$ flavors \footnotemark[1]
of chiral superfields
$Q_{\alpha\,I}$
\footnotemark[2]
in the fundamental representation \cite{SP}.
\footnotetext[1]
{
 The terminology used here obeys that in Ref. \cite{SP}.
 In particular ``$N_f$ flavors'' implies
that $2 N_f$ number \cite{Witten} of chiral supermultiplets
in the fundamental representation
(its dimension is $2 N_c$) under ${\rm SP}(N_c)$.
%
}
\footnotetext[2]
{
 Here $\alpha = 1, \cdots, 2 N_c$ denotes the index
of the fundamental representation of ${\rm SP}(N_c)$
and $I = 1, \cdots, 2 N_f$
are the index of the fundamental the (approximate) global
symmetry ${\rm SU}(2 N_f)$.
}
 Such a theory gives the confinement of color
but does not yield spontaneous breakdown of
chiral symmetry ${\rm SU}(2N_f)$.
 The massless degrees of freedom
appropriate to describe the infrared aspects
are the gauge singlet meson chiral superfield
$V_{IJ}$ $\sim$ $Q_{\alpha\,I} Q_{\beta\,J} {\cal J}^{\alpha\beta}$,
where ${\cal J}^{\alpha\beta}$
is an SP($N_c$)-invariant second-rank antisymmetric tensor,
transforming as a second-rank antisymmetric tensor
under ${\rm SU}(2N_f)$.
 The gauge theory induces the nonperturbative term
to the superpotential which are expressed in terms of
this meson field
\begin{eqnarray}
 {\cal W}_{\rm dyn} &=& {\rm Pf}_{N_f}(V) \\
 &\equiv&
 \displaystyle{
  \frac{1}{2^{N_f} N_f ! \,\Lambda^{2N_f - 3}}
  \epsilon^{I_1 I_2 \cdots I_{2 N_f - 1} I_{2 N_f}}
  V_{I_1\,I_2} \cdots V_{I_{2 N_f - 1}\,I_{2 N_f}}
 },
\end{eqnarray}
where $\Lambda$ is
the dynamical scale of ${\rm SP}(N_c)$ gauge theory
in some renormalization scheme.
 The aspect of the infrared dynamics
of a series of this gauge theory
has resemblance to the SU(2) gauge theory with three flavors
(six chiral supermultiplets of spinorial representation),
which plays the central role in the models of Ref. \cite{NS}.

 Hereafter we will concentrate on the case
of $N_c = 2$ which will turn out to be the minimal choice
for the purpose to illustrate
how the hierarchy in the yukawa coupling constants arises
between the two different generations of up-type quarks
( which will be identified as top and charm quarks ).
 The three generation model can be constructed by using
${\rm SP(3)}$ $\times{\rm SP(2)}$ $\times{\rm SU(2)_T}$
as confining gauge forces.

 The gauge group of the toy model
is G = ${\rm SP}(2) \times {\rm SU(2)_T} \times {\rm G_{ST}}$,
where ${\rm G_{ST}}$ =
${\rm SU(3)_C} \times {\rm SU(2)_L} \times {\rm U(1)_Y} $
is the gauge group of the standard model.
 The additional ${\rm SU(2)_T}$
is the confining gauge force
making the top quarks and the Higgs doubles.
 Table \ref{tab:sp-particle} lists
the content of the particles which
are non-singlets under SP(2) and their charges under G.

 The gauge anomaly cancellation with respect to ${\rm G_{ST}}$
is realized by adding the particles with the charges shown in 
Table \ref{tab:singlet}.
 Note that the number of each nonzero ${\rm G_{SM}}$-charge
in SP(2) non-singlet sector
is the same as those of the two generations of Nelson-Strassler model
\cite{NS}.
 In order to suppress the baryon number (B) and/or the lepton number
violating interactions,
the ${\rm U(1)_B}$ conservation and $Z_2$ symmetry are assumed
to hold \cite{NS} for the charge assignment listed in Table
\ref{tab:sp-particle} and \ref{tab:singlet}.
 The fermionic coordinate $\theta$ ($\bar{\theta}$)
is also charged under $Z_2$ in such a way that
$\theta^2$ ($\bar{\theta}^2$) has odd parity \footnotemark[5] .
\footnotetext[5]
{
 Precisely $\theta$ ($\bar{\theta}$) has the $Z_4$-charge $\omega$
($\omega^3$) with $\omega = e^{i\pi/2}$
and the imposed symmetry should be called as $Z_4$,
but this is not important for the present argument.
 Note that the gaugino has charge $+1$ so that
the soft supersymmetry breaking gaugino mass term
is not forbidden by this symmetry.
}
 Thus $Z_2$-charge of each term
in the superpotential is required to be odd.
 The superpotential
invariant under ${\rm G}\times {\rm U(1)_B} \times Z_2$
for the renormalizable terms
containing only the particles
in Table \ref{tab:sp-particle} and \ref{tab:singlet} is
\begin{eqnarray}
 {\cal W}_{\rm ren-1} &=&
 \displaystyle{
   \lambda_H\,\bar{H}_2 [{\bf h}{\bf n}]
   + \lambda_D\,\bar{D}_2 [{\bf d}{\bf n}]
   + \frac{1}{2} \lambda_E\,E^{(2)} [{\bf h}{\bf h}]
 } \nonumber \\
 && \ 
 \displaystyle{
   + \lambda_S \left<[{\bf s}{\bf s}]\right> T
   + \lambda_T\,T^3
   + \sum_{f,g=2}^3 y_L^{fg}
       \bar{e}_{R\,f} l^f \bar{H}_g
 },
 \label{eq:ren}
\end{eqnarray}
where each $[,\,]$ implies the contraction of SP(2) indices
through ${\cal J}^{\alpha\beta}$,
while ${\rm SU(2)_T}$ indices are contracted
within $<,\,>$.
 All the coefficients except for the lepton yukawa coupling constants
are ${\cal O}(1)$.
 This should be understood for any tree level superpotential
terms hereafter.

 For the completion of the toy model,
we have to care
about the generation of masses for bottom and strange quarks.
 Unfortunately the present toy model has a disaster
on this point.
 Here we would like to content ourselves
with merely presenting a self-contained model
and leave it as a future subject.

 The renormalizable superpotential
each term of which contains at least one of
the particles in Table \ref{tab:bottom-mass} is as follows;
\begin{eqnarray}
 {\cal W}_{\rm ren-2} &=&
 \displaystyle{
   M_N \left< \underline{\bar{N}}\,\underline{N} \right>
   + M_P\,[{\bf \bar{P}} {\bf P}]
  + \tau_{\bar{N}}
     \left< \underline{\bar{N}} [{\bf P}{\bf s}] \right>
  + \tau_N
     \left< \underline{N} [{\bf \bar{P}}{\bf s}] \right>
 } \nonumber \\
 &&
 \displaystyle{
  + \sum_{f=2}^3 
    \left\{
      \tau_H^f \bar{H}_f [{\bf P}{\bf h}]
      +
      \tau_D^f \bar{D}_f [{\bf P}{\bf d}]
      +
      \tau_{\bar{d}_R}^f \bar{d}_{R\,f}\,[{\bf \bar{P}}{\bf d}]
    \right\}
 }.
 \label{eq:ren2}
\end{eqnarray}
 The mass $M_P$ is taken as larger than $\Lambda_{\rm SP}$,
while $M_N$ should be larger than ${\rm \Lambda_T^h}$,
the dynamical scale of ${\rm SU(2)_T}$ gauge theory
with four spinors.
 Here it is also assumed to be less than $\Lambda_{\rm SP}$.

 Below the scale $M_P$, ${\bf P}$ and ${\bf \bar{P}}$
decouple from the theory.
 Integrating out those modes in (\ref{eq:ren2})
leads
\begin{eqnarray}
 {\cal W}_{\rm ren-2} &=&
 \displaystyle{
   M_N \left< \underline{\bar{N}}\,\underline{N} \right>   
 } \nonumber \\
 &&
 \displaystyle{
  - \frac{1}{M_P}
    \left\{
     \tau_N \tau_{\bar{N}}
       \left[
        \left< \underline{\bar{N}}{\bf s} \right>
        \left< {\bf s} \underline{N} \right>
       \right]
     +
     \tau_{\bar{N}}
     \sum_{f=2}^3 \tau_{\bar{d}_R}^f
       \bar{d}_{R\,f}
        \left< [{\bf d}{\bf s}] \underline{\bar{N}} \right>
    \right.
 } \nonumber \\
 &&
 \displaystyle{
  \left.
    + \tau_N
    \sum_{f=2}^3 \tau_H^f \bar{H}_f
     \left< [{\bf h}{\bf s}] \underline{N} \right>
    + \sum_{f,g=2}^3 \tau_{\bar{d}_R}^f \tau_H^g
      \bar{d}_{R\,f} [{\bf d} {\bf h}] \bar{H}_g
   \right\}
 } , \nonumber \\
 &&
 + \cdots , 
 \label{eq:ren2-2}
\end{eqnarray}
where the dots denote the additional
terms including $D$ or $\bar{E}$
which are not interested here.

 When the energy scale of the system
decreases further, SP(2) gauge interaction becomes strong.
 Then the gauge interactions of G other than SP(2)
and the superpotential terms
are considered as small
compared to the gauge coupling constant of SP(2)
above and around the dynamical scale $\Lambda_{\rm SP}$
so that they can be ignored approximately.
 Thus the theory is
nothing but the supersymmetric SP(2) gauge theory
with four flavors of ``quarks'' $Q_{\alpha I}$ =
$({\bf s}, {\bf d}, {\bf h}, {\bf n})_\alpha$
as promised.

 The massless composite degrees of freedom
consist of the preons
$\underline{d}$, $\underline{h}$ and $\underline{n}$
which appear in the models by Nelson and Strassler \cite{NS},
the singlet $S$,
and the composite chiral supermultiplets
$\tilde{V}_{\hat{I}\hat{J}}$ corresponding to
the second generation of Nelson-Strassler model
with the indices $\hat{I}$ and $\hat{J}$
attached to SU(6) subgroup
of the total chiral symmetry SU(8).
 Explicitly $\tilde{V}_{\hat{I}\hat{J}}$ is expressed as
\begin{equation}
 \tilde{V}_{\hat{I}\hat{J}} \sim
 \Lambda_{\rm SP}
 \left(
   \begin{array}{ccc}
     \bar{c}_R  & q^{(2)}_L & - D^{(2)} \\
    - q^{(2)}_L & \bar{E}_2 & - H^{(2)} \\
      D^{(2)}   & H^{(2)}   & 0
   \end{array}
 \right).
 \label{eq:matter}
\end{equation}
 There appear
the right-handed charm quark $\bar{c}_R$,
the left-handed quark doublet $q^{(2)}_L$ of the second generation,
and the Higgs doublets $H^{(2)}$,
which will not obtain nonzero vacuum expectation value (VEV)
as its mass squared will be found of the order $\Lambda_{\rm SP}^2$,
much larger than the negative corrections
by the large top yukawa coupling
of the order $ - (y_t m_{\rm susy}/(4\pi))^2$
with the supersymmetry breaking scale $m_{\rm susy}$ $\sim$
$ {\cal O}(1) $ TeV and
top yukawa coupling constant $y_t$ of order one.

 In contrast to the Nelson-Strassler model
the nonperturbatively generated superpotential terms are
now the non-renormalizable ones {\it ab initio}
\begin{equation}
 W_{\rm dyn} = {\rm Pf}_{2}(V),
\end{equation}
which explicitly give rise to
\begin{eqnarray}
 \Lambda_{{\rm SP}}\,W_{\rm dyn} &=&
 \displaystyle{
   S\,
   \left\{
      \alpha_2^\prime \,\bar{c}_R\, q^{(2)}_L H^{(2)}
    + \beta_2^\prime \, \bar{c}_R\, D^{(2)} \bar{E}_2
    + \gamma_2^\prime \, D^{(2)} q^{(2)}_L q^{(2)}_L
   \right\}
 } \nonumber \\
 &&
 \displaystyle{
   +
   \alpha_2^H\,
     \bar{c}_R\, q^{(2)}_L 
     \left<\underline{h}\,\underline{n} \right>
   +
   \alpha_2^{t_R}\,
     \left< \underline{d}\,\underline{d} \right>
     q^{(2)}_L H^{(2)}
   +
   \alpha_2^{q_L}\,
     \bar{c}_R
     \left< \underline{d}\,\underline{h} \right> H^{(2)}
 } \nonumber \\
 &&
 \displaystyle{
   +
   \beta_2^{t_R}\,
     \left< \underline{d}\, \underline{d} \right>
     D^{(2)} \bar{E}_2
   +
   \beta_2^D\,
     \bar{c}_R \left< \underline{d}\, \underline{n} \right>
     \bar{E}_2
   +
   \beta_2^E\,
     \bar{c}_R D^{(2)}
     \left< \underline{h}\, \underline{h} \right>
 } \nonumber \\
 &&
 \displaystyle{
   +
   \gamma_2^{q_L}\,
     \left< \underline{d}\, \underline{h} \right> q^{(2)}_L D^{(2)}
   +
   \gamma_2^D\,
     \left< \underline{d}\, \underline{n} \right> q^{(2)}_L q^{(2)}_L
 },
 \label{eq:np-term1}
\end{eqnarray}
where $\alpha^\prime$ etc., are ${\cal O}(1)$ constants 
parametrizing the corrections from
the perturbative couplings,
which depend on the species taking part in each above coupling.
 The ${\rm U(1)_B}$ is conserved by
the dynamically generated terms
since they do not break chiral symmetry SU(8)
and the baryon number restricted to SP(2) non-singlet sector
is one of the generators of SU(8).
 The tree level superpotential in (\ref{eq:ren})
now turns into
\begin{eqnarray}
 {\cal W}_{\rm ren-1} &=&
 \displaystyle{
   \lambda_H \Lambda_{\rm SP} \,\bar{H}_2 H^{(2)}
   + \lambda_D \Lambda_{\rm SP}\,\bar{D}_2 D^{(2)}
   + \lambda_E \Lambda_{\rm SP}\,E^{(2)} \bar{E}_2
   + \lambda_S \Lambda_{\rm SP}\,S T
 } \nonumber \\
 && \ 
 \displaystyle{
   + \lambda_T\,T^3
   + \sum_{f,g=2}^3 y_L^{fg}
       \bar{e}_{R\,f} l^f \bar{H}_g
 }.
 \label{eq:ren-confine}
\end{eqnarray}
 Eq. (\ref{eq:ren2-2}) also becomes
\begin{eqnarray}
 {\cal W}_{\rm ren-2} &=&
 \displaystyle{
   M_N \left< \underline{\bar{N}}\, \underline{N} \right>   
 } \nonumber \\
 &&
 \displaystyle{
   - \frac{\Lambda_{\rm SP}}{M_P}
     \left\{
     \sum_{f=2}^3
     \left(
       \tau_{\bar{N}}
       \tau_{\bar{d}_R}^f
         \bar{d}_{R\,f}
         \left< \underline{d}\, \underline{\bar{N}} \right>
       +
       \tau_N \tau_H^f \bar{H}_f
       \left< \underline{h}\, \underline{N} \right>
     \right)
    + \sum_{f,g=2}^3 \tau_{\bar{d}_R}^f
       \tau_H^g\,
       \bar{d}_{R\,f} q^{(2)}_L \bar{H}_g
  \right\}
 } \nonumber \\
 &&
 + \cdots.
 \label{eq:tree3}
\end{eqnarray}
 Now $H^{(2)}$, $\bar{H}_2$,
$D^{(2)}$, $\bar{D}_2$, $\bar{E}^{(2)}$, $E_2$, $S$ and $T$
gain masses of order ${\rm \Lambda_{SP}}$
and decouple from the effective theory under those thresholds,
producing new non-renormalizable operators,
but irrelevant for mass generation.
 Keeping the relevant terms for the quark yukawa couplings explicitly
in the sum of (\ref{eq:np-term1}), (\ref{eq:ren-confine})
and (\ref{eq:tree3})
the effective superpotential becomes
\begin{eqnarray}
 {\cal W}_{\rm eff} &=&
 \displaystyle{
   \alpha_2^H \frac{1}{\Lambda_{\rm SP}}\,
     \bar{c}_R\, q^{(2)}_L
     \left< \underline{h}\, \underline{n} \right>
 } \nonumber \\
 &&
 \displaystyle{
  - \frac{\Lambda_{\rm SP}}{M_P}
    \left\{
     \tau_{\bar{N}}
     \sum_{f=2}^3 \tau_{\bar{d}_R}^f
       \bar{d}_{R\,f}
       \left< \underline{d}\, \underline{\bar{N}} \right>
    + \tau_N \tau_H
       \bar{H} \left< \underline{h}\, \underline{N} \right>
    + \tau_H
    \sum_{f=2}^3 \tau_{\bar{d}_R}^f
        \bar{d}_{R\,f} q^{(2)}_L \bar{H}
   \right\}
 } \nonumber \\
 &&
 \displaystyle{
   + \cdots
 },
\end{eqnarray}
by writing $\bar{H}$ for $\bar{H}^{(3)}$ .
	
 As the energy further goes down,
$\underline{N}$, $\underline{\bar{N}}$ decouples
and the theory becomes described by the effective superpotential
\begin{eqnarray}
 {\cal W}_{\rm eff} &=&
 \displaystyle{
   \alpha_2^H \frac{1}{\Lambda_{\rm SP}}\,
     \bar{c}_R\, q^{(2)}_L
     \left< \underline{h}\,\, \underline{n} \right>
   - \tau_H \frac{\Lambda_{\rm SP}}{M_P}
    \sum_{f=2}^3 \tau_{\bar{d}_R}^f
        \bar{d}_{R\,f} q^{(2)}_L \bar{H}
 } \nonumber \\
 &&
 \displaystyle{
  - \frac{\Lambda_{\rm SP}}{M_P\,M_N}
    \left( \tau_{\bar{N}} \tau_N \tau_H \right)
    \sum_{f=2}^3 \tau_{\bar{d}_R}^f
      \bar{d}_{R\,f}
      \left< \underline{d}\, \underline{h} \right> \bar{H}
  + \cdots
 }.
 \label{eq:s-pot}
\end{eqnarray}

 Finally the gauge force ${\rm SU(2)_T}$
becomes strong at the scale $\Lambda_{\rm T}$
enough to confine the composite preons $\underline{d}$,
$\underline{h}$ and $\underline{n}$,
where $\Lambda_{\rm T}$ is now the dynamical scale of
${\rm SU(2)_T}$ gauge theory with six spinors.
 This is the dynamical feature
due to the supersymmetric SU(2) gauge theory
with six spinors.
 The massless composite degrees of freedom
are then
\begin{equation}
 \left(
   \begin{array}{ccc}
    \left< \underline{d}\, \underline{d} \right> &
    \left< \underline{d}\, \underline{h} \right> &
    \left< \underline{d}\, \underline{n} \right> \\
    \left< \underline{h}\, \underline{d} \right> &
    \left< \underline{h}\, \underline{h} \right> &
    \left< \underline{h}\, \underline{n} \right> \\
    \left< \underline{n}\, \underline{d} \right> &
    \left< \underline{n}\, \underline{h} \right> &
    0
   \end{array}
 \right)
 \sim {\rm \Lambda_{T}}
 \left(
   \begin{array}{ccc}
     \bar{t}_R & q^{(3)}_L & -D^{(3)} \\
     - q^{(3)}_L & \bar{E}_3 & - H \\
     D^{(3)} & H & 0
   \end{array}
 \right).
\end{equation}
 ${\rm SU(2)_T}$ gauge dynamics generates
the additional new cubic terms
to the superpotential (\ref{eq:s-pot}) nonperturbatively
\begin{eqnarray}
 {\cal W}_{\rm eff} &=&
 \displaystyle{
  \alpha\,\bar{t}_R\,q^{(3)}_L H
  + \alpha_2^H \frac{\Lambda_{\rm T}}{\Lambda_{\rm SP}}\,
     \bar{c}_R\,q^{(2)}_L H
   - \tau_H \frac{\Lambda_{\rm SP}}{M_P}
    \sum_{f=2}^3 \tau_{\bar{d}_R}^f
        \bar{d}_{R\,f} q^{(2)}_L \bar{H}
 } \nonumber \\
 &&
 \displaystyle{
   - \frac{\Lambda_{\rm SP} \Lambda_{\rm T}}{M_P\,M_N}
     \left( \tau_{\bar{N}} \tau_N \tau_H \right)
     \sum_{f=2}^3 \tau_{\bar{d}_R}^f
       \bar{d}_{R\,f} q^{(3)}_L \bar{H}
   +  \cdots
 }.
 \label{eq:eff}
\end{eqnarray}

 Eq. (\ref{eq:eff}) shows that
the yukawa coupling constant for top quark
appears as ${\cal O}(1)$ \cite{Strassler,NS}
while that for charm quark is damped by the factor
$\Lambda_{\rm T}/\Lambda_{\rm SP}$, the ratio of
the confining scales of the two gauge dynamics.
 Note that the origin of this factor
arises from the non-renormalizable nature
of ${\rm SP}(N_c)$ dynamics and
the multi-compositeness of the Higgs doublet
acquiring nonzero VEV.
 However as one linear combination of $\bar{d}_R$
participates in (\ref{eq:eff}),
only one down-type quark gains the mass,
which is strange quark as
the yukawa coupling in matter involves
$q^{(2)}_L$.
 This unpleasant feature in the present example
originates
from the multi-compositeness of $q^{(3)}_L$
which prevents a renormalizable superpotential term
and leaves a global symmetry in the yukawa-sector.
 Such a pattern that
bottom becomes lighter than strange quark
will not be circumvented
as long as these quarks are considered as elementary objects
since the dimension of the operator
containing the third-generation quark doublet $q^{(3)}_L$
necessarily becomes higher than that for $q^{(2)}_L$
when they are written down
in terms of their most fundamental objects.
 Obviously such a feature originates
from our present insight on the structure of matters.
 However compositeness of the right-handed down-type quarks
will turn over this invalid relations
if confinement producing bottom quark
occurs at much higher scale than that making strange quark,
or strange quark has deeper substructure
than bottom quark.

\section{Discussion and conclusion}
\label{sec:conclusion}

 Here we seek the possibility
that the mass hierarchy of the matters
is directly related with
the compositeness hierarchy of the matters.
 The toy model is given to
illustrate
that the dynamically generated nonrenormalizable
terms induces the suppression factor
for the light up-type quark yukawa coupling.
 This factor is the ratio of the two confining scales,
differing from that in Ref. \cite{Strassler}
using similar ${\rm SP}(N_c)$ gauge dynamics,
in which the yukawa coupling for the light flavors
arises from the non-renormalizable terms
expected to be induced from Planck scale physics.
 The compositeness of Higgs doublet $H$ also plays
an essential role here.

 It was also observed that
the down-type quarks must be considered
as composite to reach the realistic pattern of their masses
once we follow up the present idea.
 Accepting this fact positively,
this requirement
naturally suggests
that all the matters should be composite
if we follow the present approach.

 The multi-compositeness of the matter
often forbids the renormalizable superpotential at tree level
assuring an accidental symmetry.
 Such an inherent property might become useful
to generate desirable hierarchy in the cleverer model,
although it yielded some unpleasant
results ( including $\mu$ problem as discussed below )
in the toy model here.

 The exemplifying model does not generate the mass term
for Higgs bosons (so called $\mu$ term) as well as
the mass terms for $D^{(3)}$, $\bar{D}_3$, $E^{(3)}$ and $\bar{E}_3$
from the renormalizable interactions.
 The nonrenormalizable operators to induce it
can always be put in by hand, for instance, as
\begin{eqnarray}
 \frac{1}{M_1 M_2}
 \bar{H} \left< [{\bf h}{\bf s}][{\bf s}{\bf n}] \right>,
\end{eqnarray}
for the Higgs mass terms,
but it is rather unattractive
as the mass scale entering above
must be tuned to give a right order of magnitude
as ${\rm \Lambda_{SP}^2 \Lambda_T}/(M_1 M_2)$
$\sim$ ${\cal O}$ (1) [TeV].
 This point also needs further consideration
in the construction of the realistic model
based on the present approach,
which is left over as the future subject.

\acknowledgements

 The author thanks N. Kitazawa
for discussing about the supersymmetric gauge theory.

%

%
%
\begin{table}
 \caption{SP(2) non-singlet species and their quantum numbers.
          ${\rm U(1)_B}$ is the baryon number.}
 \vspace{0.3cm}
 \begin{tabular}{cccccccc}
   & SP(2) & ${\rm SU(2)_T}$  &
     ${\rm SU(3)_C}$ & ${\rm SU(2)_L}$ & ${\rm U(1)_Y}$ &
     ${\rm U(1)_B}$ & $Z_2$ \\
   \hline
   ${\bf s}$ & {\bf 4} & {\bf 2} & {\bf 1} & {\bf 1} & 0 &
     0 & $+$ \\
   ${\bf d}$ & {\bf 4} & {\bf 1} &
    {\bf 3} & {\bf 1} & $-\frac{1}{3}$ &
     $-\frac{1}{6}$ & $+$ \\
   ${\bf h}$ & {\bf 4} & {\bf 1} &
    {\bf 1} & {\bf 2} & $\frac{1}{2}$ &
     $\frac{1}{2}$ & $+$ \\
   ${\bf n}$ & {\bf 4} & {\bf 1} &
    {\bf 1} & {\bf 1} & 0 &
     $-\frac{1}{2}$ & $-$
 \end{tabular}
 \label{tab:sp-particle}
\end{table}
\begin{table}
 \caption{SP(2) singlet species and their charges
          under symmetry group,
          in which each of them is introduced in duplicate
          ( $f = 2, 3$ ), except for $T$.}
 \vspace{0.3cm}
 \begin{tabular}{cccccccc}
   & SP(2) & ${\rm SU(2)_T}$  &
     ${\rm SU(3)_C}$ & ${\rm SU(2)_L}$ & ${\rm U(1)_Y}$ &
     ${\rm U(1)_B}$ & $Z_2$ \\
   \hline
   $\bar{d}_{R\,f}$ & {\bf 1} & {\bf 1} &
     ${\bf 3^*}$ & {\bf 1} & $\frac{1}{3}$ &
     $-\frac{1}{3}$ & $-$ \\
   $\bar{H}_f$ & {\bf 1} & {\bf 1} &
    {\bf 1} & {\bf 2} & $-\frac{1}{2}$ &
     0 & $+$ \\
   $E^f$ & {\bf 1} & {\bf 1} &
    {\bf 1} & {\bf 1} & $-1$ & $-1$ & $-$ \\
   $\bar{D}_f$ & {\bf 1} & {\bf 1} &
    ${\bf 3^*}$ & {\bf 1} & $\frac{1}{3}$ &
     $\frac{2}{3}$ & $+$ \\
   $\bar{e}_{R\,f}$ & {\bf 1} &  {\bf 1} &
     {\bf 1} & {\bf 1} & 1 & 0 & $+$ \\
   $l^f$ & {\bf 1} & {\bf 1} &
     {\bf 1} & {\bf 2} & $-\frac{1}{2}$ & 0 & $-$ \\
   $T$ & {\bf 1} & {\bf 1} & {\bf 1} & {\bf 1} &
     0 & 0 & $-$
 \end{tabular}
 \label{tab:singlet}
\end{table}
\begin{table}
 \caption{Additional species incorporated for generating
          the down-type quark masses.}
 \vspace{0.3cm}
 \begin{tabular}{cccccccc}
   & SP(2) & ${\rm SU(2)_T}$  &
     ${\rm SU(3)_C}$ & ${\rm SU(2)_L}$ & ${\rm U(1)_Y}$ &
     ${\rm U(1)_B}$ & $Z_2$ \\
   \hline
   $\underline{N}$ & {\bf 1} & {\bf 2} & {\bf 1} & {\bf 1} &
   0 & $-\frac{1}{2}$ & $-$ \\
   $\underline{\bar{N}}$ & {\bf 1} & {\bf 2} & {\bf 1} & {\bf 1} &
   0 & $\frac{1}{2}$ & + \\
   ${\bf P}$ & {\bf 4} & {\bf 1} & {\bf 1} & {\bf 1} &
   0 & $-\frac{1}{2}$ & $-$ \\
   ${\bf \bar{P}}$ & {\bf 4} & {\bf 1} & {\bf 1} & {\bf 1} &
   0 & $\frac{1}{2}$ & +
 \end{tabular}
 \label{tab:bottom-mass}
\end{table}

\begin{references}
%
\bibitem{Strassler}
 M. J. Strassler,
  Phys. Lett. {\bf B376} (1996) 119.
%
\bibitem{NS}
 A. E. Nelson and M. J. Strassler,
  UW-PT-96-09, hep-ph/9607362 (1996).
%
\bibitem{LM}
 M. A. Luty and R. N. Mohapatra,
  UMD-PP-97-43, hep-ph/9611343 (1996). 
%
\bibitem{Seiberg}
 N. Seiberg,
  Phys. Lett. {\bf B318} (1993) 469;
  Phys. Rev. D {\bf 49} (1994) 6857; 
 K. Intriligator, R. G. Leigh and N. Seiberg,
  Phys. Rev. D {\bf 50} (1994) 1092;
 K. Intriligator and N. Seiberg,
  Nucl. Phys. Proc. Suppl. {\bf 45BC} (1996) 1. 
%
\bibitem{Csaki}
 C. Cs{\'a}ki, M. Schmaltz and W. Skiba,
  MIT-CTP-2597, BUHEP-96-46,
  hep-th/9612207 (1996).
%
\bibitem{SP}
 K. Intriligator and P. Pouliot,
  Phys. Lett. {\bf B353} (1995) 471.
%
\bibitem{Witten}
 E. Witten,
  Phys. Lett. {\bf B117} (1982) 324.
%
\end{references}
\end{document}